\documentclass{article}

\usepackage{graphicx}
\usepackage{amsmath}
\usepackage{wasysym}
\usepackage{caption}
\usepackage{subcaption}
\usepackage{comment}
\usepackage{authblk}
\makeatletter
\renewcommand\@biblabel[1]{}
\makeatother

\title{Constraining the effect of convective inhibition on the thermal evolution of Uranus and Neptune}
\author[1]{Steve Markham\footnote{smarkham@caltech.edu}}
\author[1]{Dave Stevenson}
\affil[1]{California Institute of Technology, Dept. of Geological and Planetary Sciences}
\date{\today}
\newcommand{\Tint}{T_{\text{int}}}
\newcommand{\qmax}{q_{\text{max}}}
\newcommand{\qcrit}{q_{\text{crit}}}

\begin{document}

\maketitle

\begin{abstract}
The internal heat flows of both Uranus and Neptune remain major outstanding problems in planetary science. 
Uranus' surprisingly cold effective temperature is inconsistent with adiabatic thermal evolution models, while Neptune's substantial internal heat flow is twice its received insolation. 
In this work we constrain the magnitude of influence condensation, including latent heat and inhibition of convection, can have on the thermal evolution of these bodies. 
We find that while the effect can be significant, it is insufficient to solve the Uranus faintness problem on its own. 
Self-consistently considering the effects of both latent heat release and stable stratification, methane condensation can speed up the cool down time of Uranus and Neptune by no more than 15\%, assuming 5\% molar methane abundance. 
Water condensation works in the opposite direction; water condensation can slow down the cool down timescale of Uranus and Neptune by no more than 15\% assuming 12\% molar water abundance. 
We also constrain the meteorological implications of convective inhibition. 
We demonstrate that sufficiently abundant condensates will relax to a state of radiative-convective equilibrium requiring finite activation energy to disrupt. 
We also comment on the importance of considering convective inhibition when modeling planetary interiors. 
\end{abstract}

\section{Introduction}
Giant planet atmospheres are primarily heated by a combination of sunlight and internal heat leftover from formation. 
All giant planets except Uranus are observed to emit more infrared radiation into space than the absorbed sunlight, by approximately a factor of two. 
Jupiter's present day luminosity can be approximately explained by convective cooling from an initially hot state over the age of the solar system \cite{hubbard1977}\cite{hubbard+1999}. 
To accurately reproduce Saturn's present day state, one may need to account for additional heating by the settling of helium rain from the envelope into the interior \cite{hubbard+1999}\cite{stevenson1983}. 
However, luminosity is a crude indicator of thermal evolution since planets can store heat internally and may have internal heat sources (e.g. differentiation). 
The present luminosities of Uranus and Neptune are not well understood because even their basic structures, including composition, internal structure, and thermal transport properties, are not well understood. \\

Measurements of the ice giants' electromagnetic emission to space began in the 1960s \cite{kellerman-toth1966}, with high quality far infrared measurements constraining the effective temperatures beginning in the 1970s \cite{fazio+1976}\cite{loewenstein+1977}\cite{stier+1978}.
These early observations concluded that Uranus appeared approximately in equilibrium with its received sunlight, while Neptune emitted more than twice the radiation it received. 
These observations were corroborated by higher quality analysis after the Voyager 2 flybys \cite{pearl+1990}\cite{pearl-conrath1991}. 
The 1$\sigma$ upper limit for Uranus' energy balance (the ratio between its emitted and absorbed thermal flux) is 1.14. 
The lower limit is below unity, indicating the results are consistent with zero internal heat flow. 
However, we know the heat flow cannot be zero because Uranus has a magnetic field. 
Moreover, the higher microwave temperatures at long wavelengths (e.g. \cite{gulkis+1983}) are compatible with heat flow from depth. 
Uranus \textit{must} be convective at depth. 
\\

Theoretical attempts to explain these observations began promptly. 
It was immediately clear that the ice giants could not have the same thermal histories as the gas giants. 
Early studies concluded that, if these planets cool convectively like the gas giants, they must have formed at a temperature not much warmer than their current states \cite{hubbard1978}\cite{hubbard-macfarlane1980}, a highly unlikely interpretation because the energy of accretion $\sim G M^2/R$ far exceeds their current heat content for any plausible assumption of structure. 
Alternative theories suggested a large fraction of gravitational heat of formation remains trapped in the interior, but by some mechanism cannot escape to space \cite{podolak+1991}. 
More recent studies suggest that there is no problem for Neptune \cite{fortney+2011}\cite{linder+2019}, or even that Neptune's present luminosity is higher than expected \cite{nettelmann+2016}\cite{scheibe+2019}. 
Uranus' very low internal heat flux, sometimes known as the faintness problem \cite{helled+2020}, remains largely unsolved, although it has been suggested that the problem can be solved by modeling thin layers of static stability near phase boundaries \cite{nettelmann+2016}. 
Today it is largely accepted that the adiabatic assumption for the interior is probably inappropriate for Uranus and Neptune \cite{helled+2020}. 
\\

In this work we present a mechanism that inhibits convection near the methane cloud level, thereby trapping internal heat beneath the clouds. 
This mechanism has already been theorized and discussed e.g. \cite{leconte+2017}  \cite{friedson-gonzales2017} \cite{guillot2005}, but the effect of methane on the ice giants has not yet been explicitly quantified and worked into a thermal evolutionary model. 
In hydrogen atmospheres, sufficiently abundant condensible species can shut off convection near the cloud level \cite{guillot1995}\cite{guillot2005}. 
By ``sufficiently abundant'' we mean greater than an analytically calculable critical mole fraction $\qcrit$. 
This value is about 1.4\% for methane and 1.2\% for water under the relevant conditions in Uranus and Neptune. 
Recent theoretical study confirms this effect is also stable against double diffusive convection in a saturated medium in the fast precipitation limit, indicating radiation would be the only remaining efficient thermal transport mechanism \cite{leconte+2017} \cite{friedson-gonzales2017}. 
Methane is certainly sufficiently abundant for convective inhibition to occur \cite{helled+2020}. 
The long-term survival of the configuration against entrainment is still a subject of research; the configuration may be intermittently eroded, destroyed, and reformed \cite{friedson-gonzales2017}. 
 \\

In Section~\ref{preliminaries} we begin by laying the heuristic groundwork and providing analytic order of magnitude estimates of the effect of convective inhibition.  
Then in Section~\ref{atmosphere} we outline a more detailed atmospheric model.  We then use this model to quantify the effects of condensation on the difference between the planet's observed effective temperature and its internal entropy. 
We also comment on the meteorological implications arising from convective inhibition on Uranus, Neptune and Saturn, as well as its importance for interior modeling. 
In Section~\ref{evolution} we constrain the importance of both methane and water condensation on the planets' thermal evolution.  
Finally in Section~\ref{discussion} we make recommendations for future missions to the ice giants, and comment on additional applications of this mechanism to the thermal histories of exoplanets, especially super-Earths.

\section{Intuition and analytic approximations}
\label{preliminaries}
A hydrogen atmosphere becomes stable against convection at a critical value of the condensate mole fraction $q_{\text{crit}}$ that depends on temperature and the properties of the condensate and the gas mixture \cite{guillot1995} \cite{guillot2005} \cite{li-ingersoll2015} \cite{leconte+2017} \cite{friedson-gonzales2017}. 
The mechanism is as follows: consider an isobaric open system hydrogen gas parcel saturated with a vapor species of higher molecular weight.
Assume there exists a finite reservoir of liquid condensate in equilibrium with the saturated parcel, outside but in contact with the system. 
If the parcel is relatively cool, the effect of the condensate will be a small correction, and the parcel will approximately behave like an ideal gas such that density decreases as temperature increases. 
However, as temperature increases at fixed pressure, the mixing ratio of the condensate likewise increases. 
Because the condensate vapor is heavier than the dry air, there comes a crossover where the Arrhenius relationship in temperature governing vapor pressure saturation overcomes the linear relationship in temperature governing mean spacing between molecules in a gas. 
After this crossover point for the system outlined above, increasing temperature actually increases the density of the parcel. 
For this reason, a hydrogen atmosphere with sufficiently abundant condensate ($q_{\text{max}} > q_{\text{crit}}$) with an internal heat source will not convect. 
It does not convect because the warmer underlying gas is Ledoux stable to convection for any temperature gradient if saturated at all levels. 
In our notation, 
\begin{equation}
\label{eq:qcrit}
q_{\text{crit}}^{-1} = \left(\frac{\mu_c L}{RT}-1\right) ( \epsilon-1)
\end{equation}
where $R$ is the ideal gas constant, $L$ is the latent heat of vaporization, $\mu_c$ is the molecular weight of the condensate, and $\epsilon \equiv \mu_c/\mu_d$ is the ratio of the condensate molecular mass to that of dry air. 
Note that our definition of $q$ is the molar mixing ratio, and is different from the quantity defined as $q$ in \cite{leconte+2017}.\\

If we neglect radiative transfer of heat, we can analytically approximate the temperature difference between the top and bottom of the stable layer. 
The bottom of the upper saturated level satisfies $q=\qcrit$, while the deep well-mixed convective atmosphere has uniform composition of the vapor satisfying $q=\qmax$. 
In the limit of no thermal transport, the stable layer will reduce to a stable interface, with an unsaturated convective level beneath a saturated convective level. 
Thus, we can approximate the top of the deep well mixed convective layer and the bottom of the upper saturated convective layer to be at the same pressure level $p$. 
We define the temperature at the bottom of the saturated convective layer to be $T_1$, and the temperature at at the top of the well-mixed convective layer to be $T_2$. 
In this case, if the saturation vapor pressure $p_s$ satisfies the Clausius-Clapeyron relationship where $q=p_s/p$, we can immediately write down 
\begin{equation}
\label{eq:approx}
\frac{1}{T_2} - \frac{1}{T_1} = \frac{R}{\mu_c L} \log\left( \frac{\qcrit}{\qmax} \right)
\end{equation}
Equation~\ref{eq:approx} turns out to be a good approximation for very deep clouds (pressure of order a hundred bars, e.g. the water cloud level in the contemporary ice giants), where the atmosphere is relatively opaque. 
This approximation is less accurate when the atmosphere is less opaque and radiative transfer is more efficient, such as the methane cloud deck of the contemporary ice giants, or the water cloud deck earlier in their thermal histories. 
Equation~\ref{eq:approx} will also always be an upper bound to the difference between $T_1$ and $T_2$ because it neglects thermal transport. 
In practice the difference between the pressure levels at the top and bottom of the stable layer play an important role. 
In Section~\ref{atmosphere}, we solve the problem of radiative convective equilibrium explicitly. \\

\begin{figure}
\centering
\includegraphics[scale=.3]{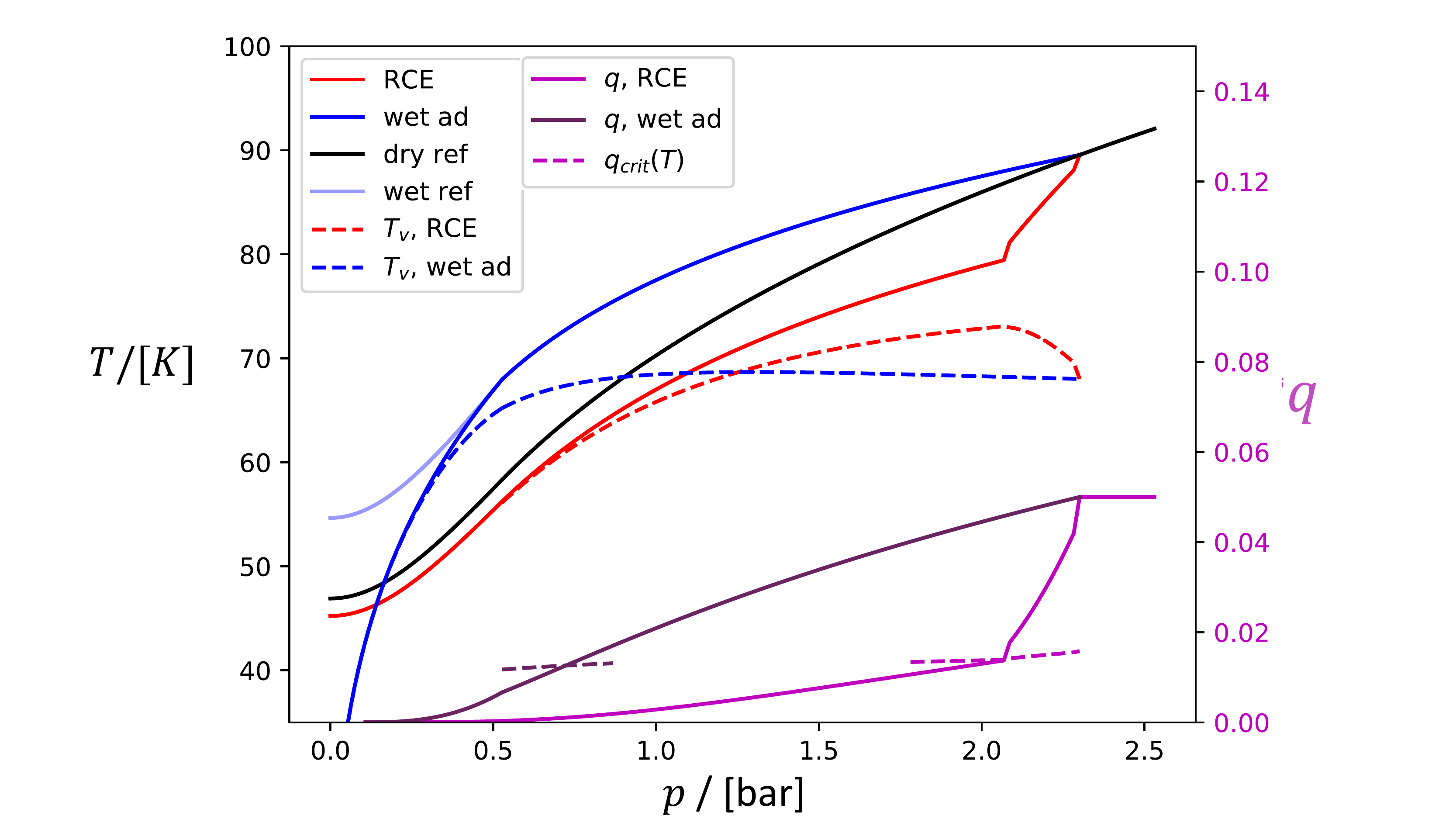}
\caption{Sample atmospheric profile useful for intuition. Blue curves show wet adiabatic (wet ad) temperature (solid) and virtual temperature (dashed) profiles. Red curves show our radiative-convective equilibrium (RCE) solutions, and the black curve shows the dry adiabatic reference solution, all corresponding to the same interior entropy.  Magenta and purple curves correspond to saturated abundance of methane for RCE and wet ad respectively, while the corresponding dashed curves show the local critical mixing ratio $\qcrit$. 
}
\label{fig:PT}
\end{figure}

The goal of this analysis is to quantify the effect of condensation on a planet's effective temperature. 
For a given internal entropy, there will be three important theoretical effective temperatures we discuss throughout this paper: $T_e$, $\Tint$, and $T_{ab}$.  
We define these quantities here, referring to Figure~\ref{fig:PT} to illustrate a sample atmospheric profile that corresponds to its respective temperature. 
The effective temperature $T_e$ (corresponding to the solid red curve on Figure~\ref{fig:PT}) accounts for convective inhibition, and should correspond to the observed effective temperature of the planet from the outside. 
The internal effective temperature $\Tint$ (faded solid blue curve on Figure~\ref{fig:PT}) accounts for latent heat but not convective inhibition.  
Finally the adiabatic effective temperature $T_{ab}$ (black curve on Figure~\ref{fig:PT}) is the effective temperature the planet would have if no condensation occurred at all, i.e. if the whole troposphere were dry adiabatic. \\

We now approximate how this temperature difference between $T_1$ and $T_2$ changes the effective temperature of the planet. 
Consider an initially convective atmosphere (for example, mixed by a cosmic ladle) that has effective temperature $\Tint$. 
If $\qmax>\qcrit$, then part of this atmosphere will be stable to convection. 
The atmosphere will cool from the top, but cannot carry that heat out convectively, causing the upper layer to relax onto a cooler adiabat until radiative-convective equilibrium is reached. 
We wish to estimate the difference between the initial effective temperature $\Tint$ and the final effective temperature $T_e$ in equilibrium. 
Entropy increment scales as $d S \propto d T/T$ in an isobaric environment, according to the first law of thermodynamics, and adiabatic processes are isentropic. 
Therefore if the difference between $T_1$ and $T_2$ is not too large, we can approximate 
\begin{equation}
\label{eq:approx2}
\frac{T_e - \Tint}{\Tint} \sim \frac{T_1-T_2}{T_2} \sim \frac{R T_2}{\mu_c L} \log\left(\frac{\qcrit}{\qmax} \right)
\end{equation}
where $T_e$ is the observed effective temperature of the planet after accounting for convective inhibition. 
This approximation neglects the non-adiabaticity due to latent heat release, which we address in the following paragraph.
\\

In reality we expect two effects arising from condensation: the tendency toward sub-adiabatic wet pseudo-adiabaticity arising from latent heat, and the tendency toward super-adiabatic stable stratification arising from convective inhibition. 
These two effects will be in opposite directions; while latent heat will tend to produce a warmer effective temperature, convective inhibition will tend to produce a cooler effective temperature with the same interior entropy. 
The former effect has been studied in detail \cite{kurosaki-ikoma2017}, while the latter effect is the subject of this work. 
We can estimate the magnitude of the latent heat effect using the definition of equivalent potential temperature that is conserved along a moist adiabat. 
\begin{equation}
\label{eq:equivpot}
\theta_e (p, T) = \theta(p,T) \exp\left[\frac{\epsilon p_s(T)}{c_p T(p-p_s(T))}\right]
\end{equation}
where $\theta\equiv T(p_0/p)^{\nabla_{ab}}$ is the potential temperature, $\nabla_{ab}\equiv \frac{\gamma-1}{\gamma}$ is the adiabatic gradient, and $\gamma$ is the Gr\"{u}neisen parameter. 
Using this, we can estimate the temperature change due to moist adiabaticity as $\frac{T_e - \Tint}{\Tint} \sim \exp \left[ \frac{\nabla_{ab} \mu_c \qmax}{R T_2} \right]$. 
Comparing this to Equation~\ref{eq:approx2} demonstrates these two quantities should not scale in the same way. 
However, they are comparable in order of magnitude under the conditions of interest. 
Therefore both effects must be accounted for explicitly in order to fully understand the effect of condensation on thermal evolution of planets with polluted hydrogen atmospheres. 
We perform this calculation in Section~\ref{atmosphere}. \\

The importance of these effects on thermal evolution are as follows. 
Because potential temperature relates linearly to a reference temperature, the temperature at all pressures will scale linearly with the temperature at some reference pressure. 
Likewise, the effective temperature of a planet scales linearly with a reference temperature in the adiabatic region, assuming constant opacity (this is actually a poor assumption, and a fully complete model must include opacity variations due to condensation explicitly.  See Section~\ref{discussion} for further details). 
Therefore, it is possible to model thermal evolution by assuming a planet's effective temperature is linearly related to its internal heat content. 
The purpose of calculating the difference between the observed effective temperature $T_e$ and the adiabatic equivalent effective temperature $T_{ab}$ is to explicitly quantify the non-linearity arising from condensation, so that thermal evolution can be modeled self-consistently. 
We carry this out in Section~\ref{evolution}.

\section{Atmospheric model}
\label{atmosphere}
We model a radiative-convective equilibrium atmosphere using a two stream gray opacity approximation for thermal radiative transfer. 
We seek to uniquely define the apparent effective temperature $T_e$ as a function of a planet's internal equivalent effective temperature $T_{\text{int}}$ and condensate abundance $\qmax$. 
Planetary and physical properties, such as surface gravity and the physical properties of the gas mixture, are considered to be fixed. \\

We also assume the planet is subject to intermittent moist convective events that overcome the potential barrier of the stable layer. 
These could occur due to instabilities caused by entrainment over long timescales \cite{friedson-gonzales2017}, rare impact events, or strong updrafts from the interior. 
The equilibrium configuration then is reached by gradual cooling, with the upper layer relaxing onto a moist pseudo-adiabat set by a different potential temperature than the adiabat that sets the interior. 
The stable layer meanwhile will have a super-adiabatic temperature gradient set by thermal radiative equilibrium, see Figure~\ref{fig:sketch}. \\

In this section we explore the effect of methane abundances varying between 2-5\%. 
The most commonly cited number for the deep mixing ratio of methane in Uranus and Neptune are 2.3\% and 2\% respectively, because these are the nominal values in the first published work on the atmospheric structure of these planets derived from radio refractivity data from Voyager \cite{lindal+1987} \cite{lindal1992}. 
However, these early works provide solutions to the refractivity data using assumptions of methane abundance between 1\% and 4\%, finding all these solutions to be theoretically compatible with the observations. 
Subsequent analysis from ground based and Hubble observations have likewise found a range of acceptable values for both planets ranging between roughly 2\%-4\% for both planets, as well as latitudinal variation in methane mixing ratio \cite{baines+1995}\cite{rages+1991} \cite{baines-smith1990} \cite{karkoschka-tomasko2011}. 
In this work we are interested in understanding how thermal evolution is affected by methane condensation. 
Since the exact mixing ratio is not precisely constrained and we are interested in this question broadly, we take sample values for methane concentration between 2-5\% to understand how the effect changes with methane abundance. 
Unsurprisingly, the effect becomes monotonically more important as concentration increases within this range, as shown in Figures~\ref{fig:methane} and \ref{fig:meth-evol}. \\

\subsection{Defining the boundaries of the stable layer}
\begin{figure}
\centering
\includegraphics[scale=.25]{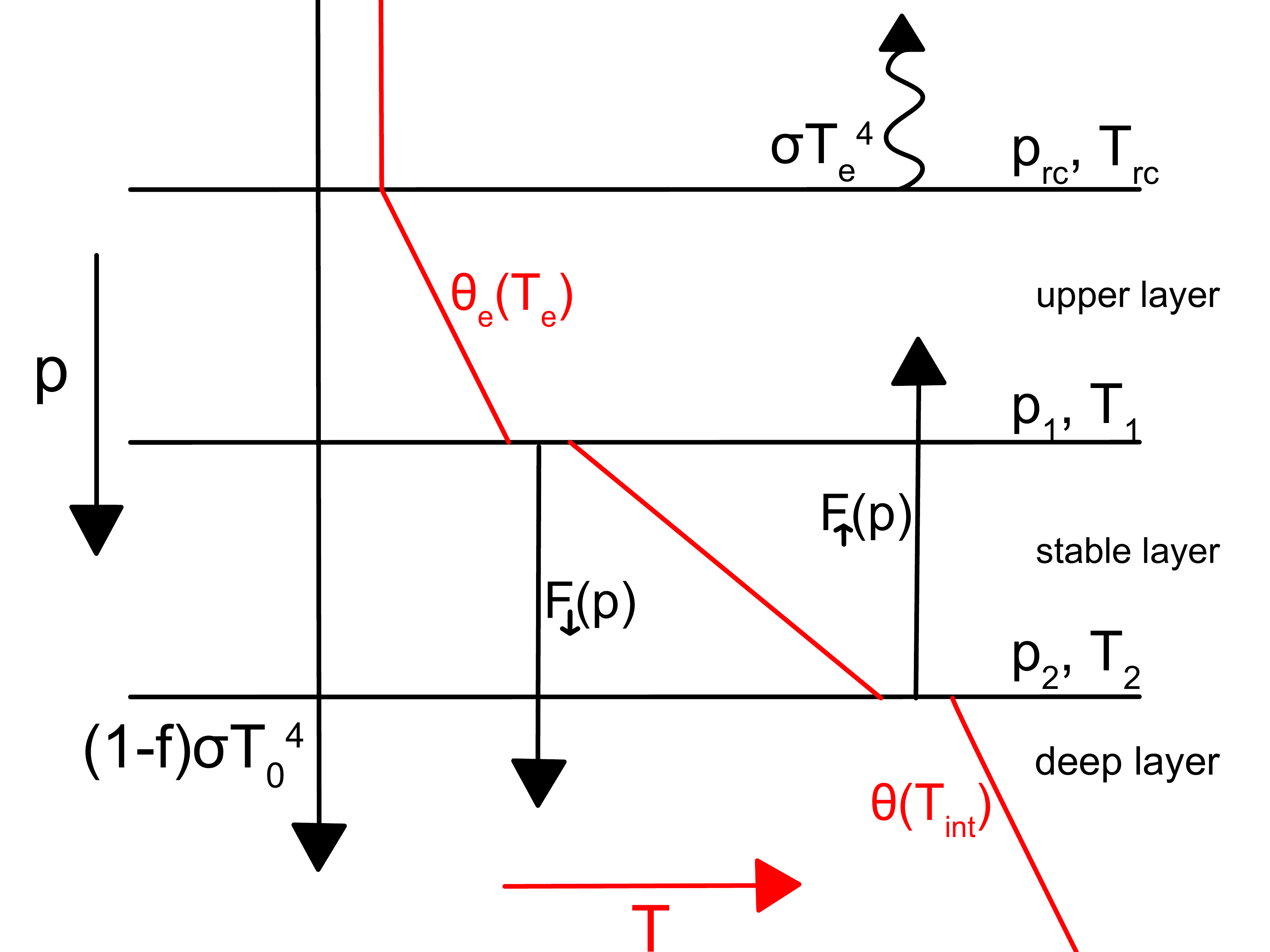}
\caption{A schematic sketch of the model (not to scale). 
The red curves represent temperature increasing to the right, while black horizontal lines demarcate important pressure levels. 
The convective layers are labeled $\theta_e(T_e)$ and $\theta(\Tint)$ respectively, to indicate that their temperature structure is dictated by setting $\theta_e$ and $\theta$ to a constant value uniquely determined by $T_e$ and $\Tint$. 
}
\label{fig:sketch}
\end{figure}

In this subsection we quantify important pressure boundaries we need to define the radiative transfer model in the following subsection. 
We are interested in the case where condensate is sufficiently abundant to inhibit convection, $\qmax > \qcrit$. 
We consider an atmosphere where optical depth unity in the IR is at lower pressure than the level at which convection is expected. 
At deeper levels (higher pressure) we assume there is a region of rapidly varying condensate mixing ratio in the vapor phase; this region can be convectively stable as discussed in Section~\ref{preliminaries}. 
Deeper still, below the conventionally defined cloud deck, the condensate mixing ratio is a constant because the vapor pressure is always less than the saturated vapor pressure at that temperature. 
We refer to this below as the ``bulk mixing ratio'' though it is strictly only applicable to whatever deep, well mixed layer lies beneath the clouds and says nothing about the actual methane abundance at far deeper levels (i.e., the methane abundance of the planet as a whole). 
Accordingly, our atmosphere has (from the top downward) a radiative layer (the stratosphere) a convective layer, another radiative layer (called the `stable layer' below) and a deep convective layer. \\

Assuming a two stream approximation with collimated light beams, the thermal structure of an atmosphere in radiative equilibrium as a function of optical depth is $T(\tau)=T_e (\tau+1/2)^{1/4}$. 
We assume the IR opacity to be dominated by pressure-induced opacity of hydrogen collisions that approximately obeys $\kappa \sim \kappa_0 (p/p_0)$, where $\kappa_0 = 10^{-2}$g$^{-1}$cm$^2$ and $p_0=1$ bar. 
Assuming a different $\kappa_0$ does not significantly affect our findings, despite altering some of the model details. 
An atmosphere in radiative equilibrium becomes unstable to convection at the point where its lapse rate becomes superadiabatic. 
Under our assumptions, this places the pressure level of the radiative-convective boundary 
\begin{equation}
p_{\text{rc}}^2 = \frac{4 g p_0 R_d}{(c_p-2 R_d)\kappa_0}
\end{equation}
where $R_d$ is the specific gas constant of dry air, and $c_p$ is its constant pressure heat capacity. 
Note for ideal gasses $R_d / c_p = \nabla_{ab}$. 
We use a dry adiabatic lapse rate to set the radiative-convective boundary because we assume the effect of moist adiabaticity is small in this relatively cold part of the atmosphere. 
Beneath the boundary, we assume the atmosphere to be moist adiabatic.
A moist adiabatic atmosphere conserves the equivalent potential temperature, Equation~\ref{eq:equivpot}. 
We set the moist adiabatic equivalent potential temperature using the temperature and pressure at the radiative-convective boundary. 
By doing this, we define a unique moist adiabat for a given effective temperatures. \\ 

In equilibrium, the mole mixing ratio is set by the condensate's saturated vapor pressure $q(p,T) = p_s(T)/p(T)$. 
When we define a moist adiabat, the temperature is uniquely defined at every pressure level. 
Therefore we can solve for the level $p_1$ where the atmosphere becomes stable to convection by solving $q(p_1) = q_{\text{crit}}(T(p_1))$, where $\qcrit$ is defined in Equation~\ref{eq:qcrit}. 
\\

Similarly, we can solve for the bottom of the stable layer by solving $q(p_2) = q_{\text{max}}$, where $q_{\text{max}}$ is the bulk abundance of the condensate species. 
This is set using the pseudoadiabat corresponding to a planet with effective temperature that neglects convective inhibition. 
By fixing the uninhibited effective temperature $\Tint$ that defines $p_2$, we can solve for the corresponding effective temperature $T_e$ that satisfied radiative convective equilibrium, i.e. $F_{\uparrow}(p_2) - F_{\downarrow}(p_2) = F_{\uparrow}(p_1) - F_{\downarrow}(p_1) = \sigma (T_e^4 - f T_0^4)$ where $f$ is the fraction of sunlight absorbed above the stable layer, and $T_0$ is the equilibrium effective temperature with the sun in the absence of internal heat (see Figure~\ref{fig:sketch}). \\

Figure~\ref{fig:sketch} contains apparent temperature discontinuities, which exist in the model. 
Of course, temperature discontinuities are not stable in natural media, as conductive heat transport will be infinite. 
Additionally, a temperature discontinuity--even if stabilized by a compositonal difference--will lead to negligible temperature differences due to thin thermal boundary layer convection in an inviscid fluid. 
The temperature discontinuities in Figure~\ref{fig:sketch} therefore do not actually represent discontinuities in nature, but steep temperature gradients. 
We provide the following order of magnitude analysis to determine how important these steep quasi-discontinuities are in the context of the model. 
The relaxation timescale for thermal diffusion in a medium with thermal conductivity $k_t$ is $\tau_c \sim \frac{\rho c_p D^2}{k_t}$, where $D$ is the vertical length scale of relevance. 
The timescale for radiative relaxation is the radiative time constant $\tau_{\text{rad}} \sim \frac{c_p}{8 \sigma T^3 \kappa}$. 
We solve for the thickness of the conductive layer by equating the two timescales solving for $D$. 
Using appropriate parameters for hydrogen around 1 bar ($k_t \sim 10^4$g~cm~s$^{-3}$~K$^{-1}$, $\rho \kappa \sim 10^{-6}$) we find the length scale to be of order 10-100 meters, small compared to atmospheric length scales (i.e., the scale height). 
Then using Fourier's Law, we find heat conduction to be of order $10^{-2}$erg~cm$^{-2}$~s$^{-1}$ for discontinuities of order 1K (scaling linearly with the size of the temperature discontinuity), about four orders of magnitude smaller than $\sigma T_e^4$ and therefore not included in the model. 
Thus the discontinuities in Figure~\ref{fig:sketch} are really there in the model, but are physically understood to be steep temperature gradients nevertheless unimportant for the purposes of calculating total heat flow. \\

We must also comment on entrainment by convection outside the stable layer. 
Entrainment will tend to erode and thin the stable layer over time \cite{friedson-gonzales2017}. 
In general for water, the erosion timescale is greater than the cooling timescale, indicating that the equilibrium configuration should exist at some times. 
However as the stable layer is eroded and becomes thinner, heat transport across the stable layer will be enhanced, reducing the difference between $T_e$ and $\Tint$. 
Eventually the thinning stable region will reduce to an interface that may be stable, unstable, or conditionally stable. 
We acknowledge that these complications are confounding factors for our model, and therefore our results that neglect entrainment erosion of the stable layer should be thought of as an upper bound on the magnitude of the effect on $\Delta T \equiv T_e - \Tint$ and on evolution. 
\\

\subsection{Radiative transfer across the stable layer}
\begin{figure}
\begin{subfigure}{.5\textwidth}
	\centering
	\includegraphics[scale=.4]{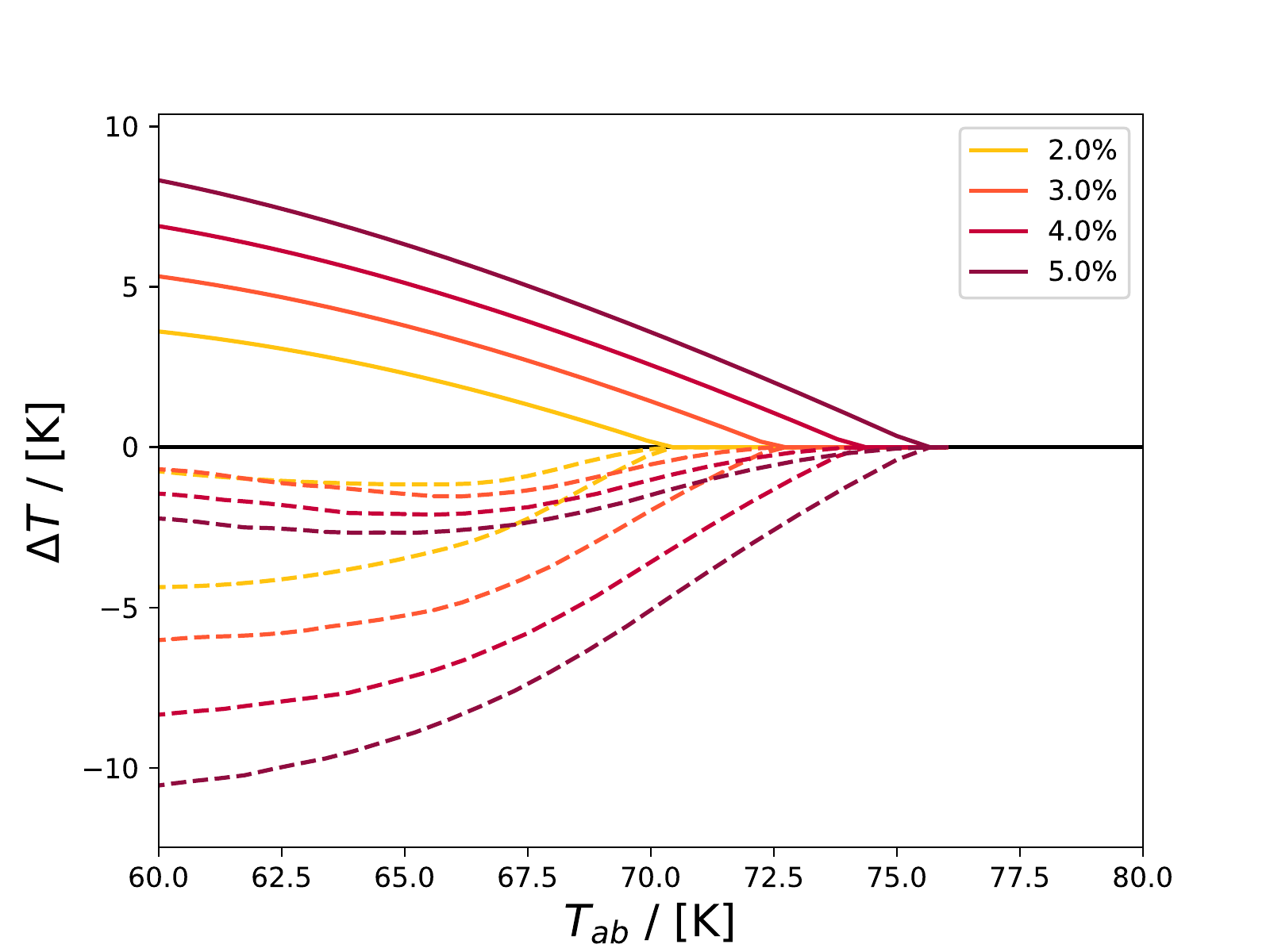}
	\caption{Effective temperature changes arising from different sources. Latent heat (solid), convective inhibition (dashed-dotted), and the net effect (dashed).}
	\label{fig:methane-cancel}
\end{subfigure}\hspace{5mm}%
\begin{subfigure}{.5\textwidth}
	\centering
	\includegraphics[scale=.4]{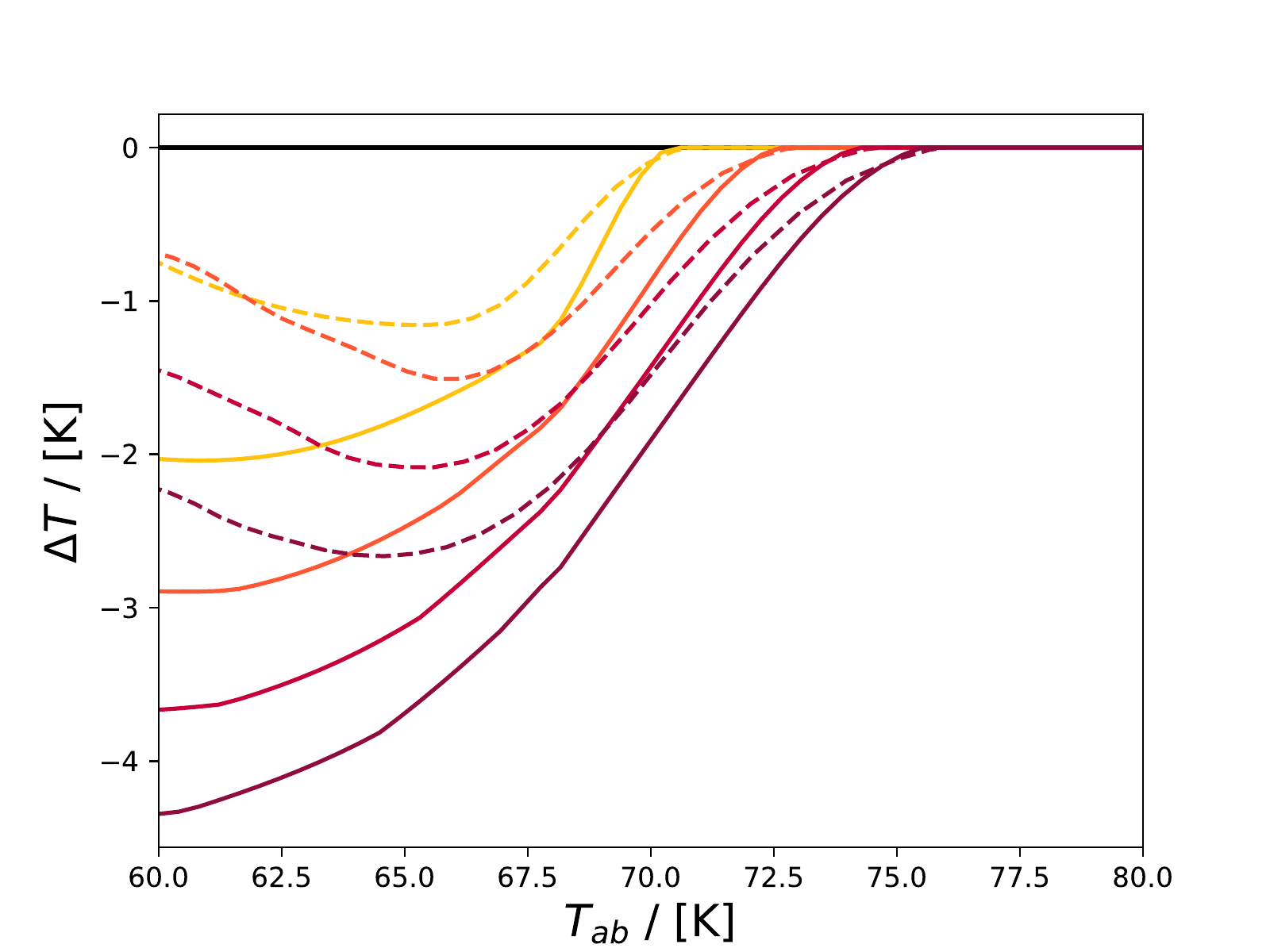}
	\caption{Net temperature differences, for an initially dry adiabatic atmosphere (solid) and an initially moist adiabatic upper layer (dashed)}
	\label{fig:nep-meth-evol}
\end{subfigure}%
\caption{$\Delta T(T_{ab})$ for different envelope abundances of methane $\qmax$ between 2-5\%, where $\Delta T \equiv T_e - T_{ab}$.  The dashed (net $\Delta T$) curves are the same in (a) and (b). 
}
\label{fig:methane}
\end{figure}

\begin{figure}
\begin{subfigure}{.5\textwidth}
	\centering
	\includegraphics[scale=.4]{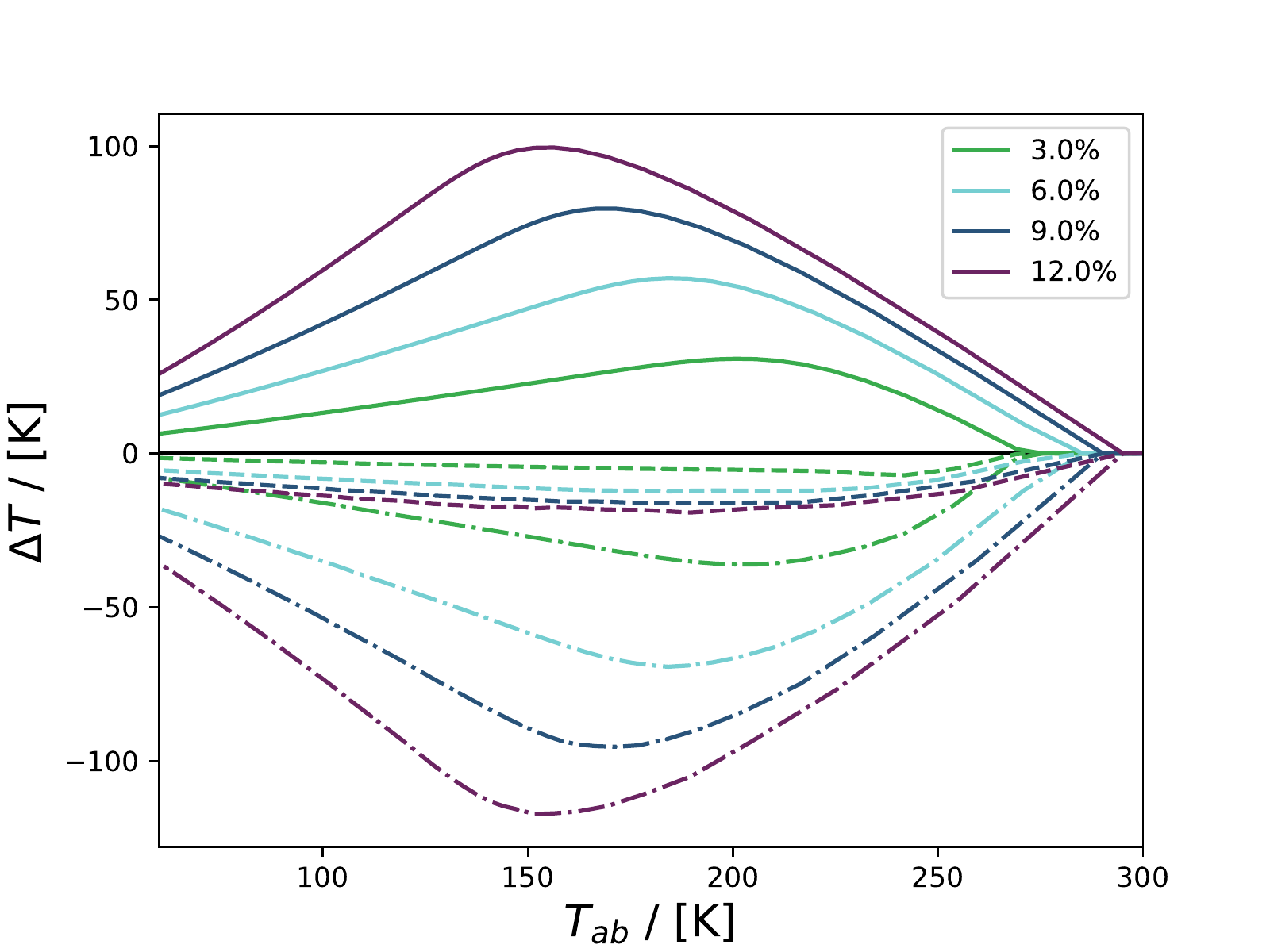}
	\caption{Effective temperature changes arising from different sources. Latent heat (solid), convective inhibition (dashed-dotted), and the net effect (dashed).}
	\label{fig:methane-cancel}
\end{subfigure}\hspace{5mm}%
\begin{subfigure}{.5\textwidth}
	\centering
	\includegraphics[scale=.4]{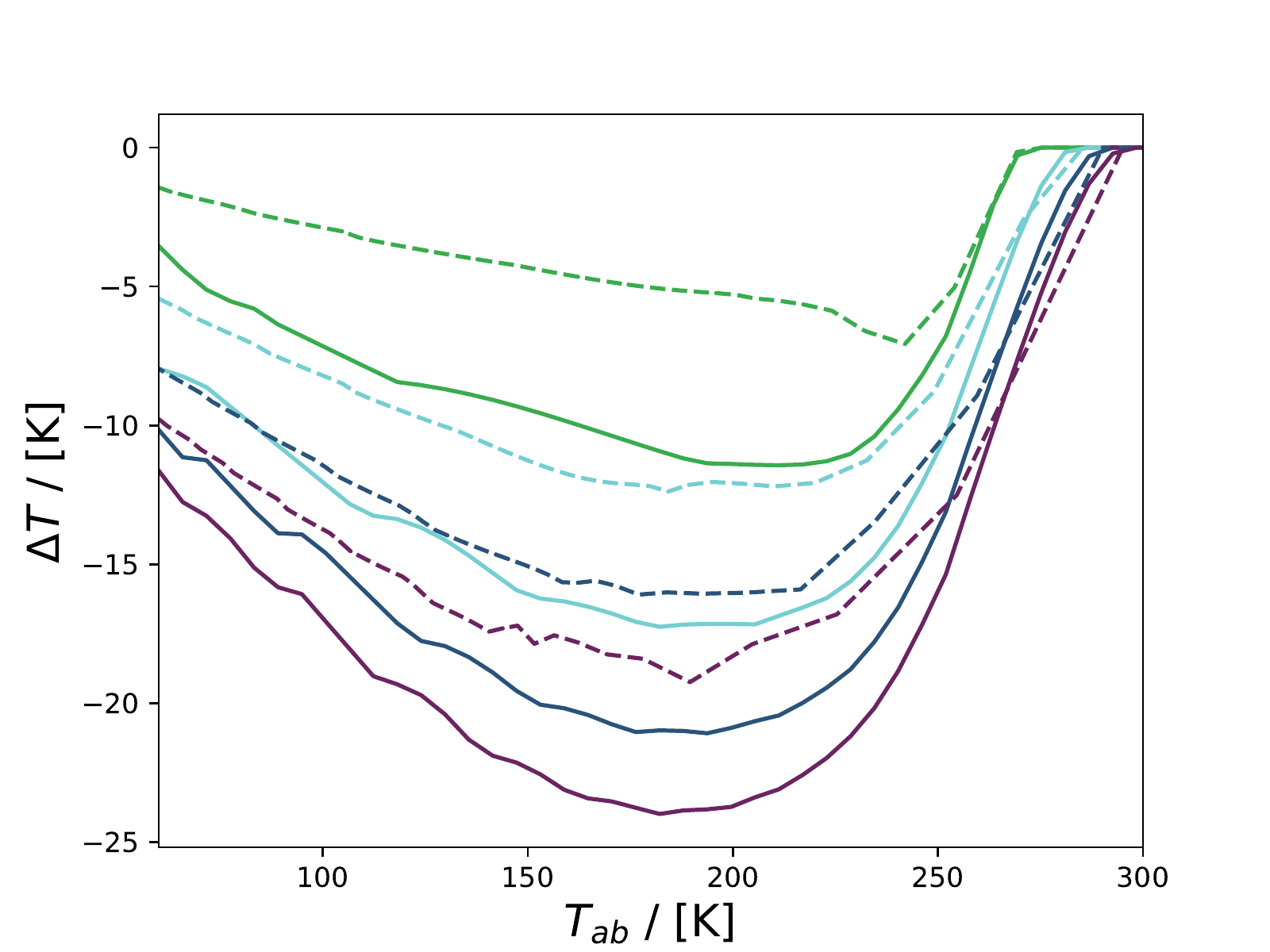}
	\caption{Net temperature differences, for an initially dry adiabatic atmosphere (solid) and an initially moist adiabatic upper layer (dashed)}
	\label{fig:nep-meth-evol}
\end{subfigure}%
\caption{$\Delta T(T_{ab})$ for different envelope abundances of water $\qmax$ between 3-12\%, where $\Delta T \equiv T_e - T_{ab}$.  The dashed (net $\Delta T$) curves are the same in (a) and (b). 
}
\label{fig:water}
\end{figure}

Figure~\ref{fig:sketch} is a useful visual reference for this section. 
In order to compute the radiative-convective equilibrium solution, we first solve for the equilibrium heat flow for a system specifying the boundaries of the stable layer $p_1$ and $p_2$, along with their corresponding temperatures $T_1$ and $T_2$. 
The temperature structure above $p_1$ is moist adiabatic, while the temperature structure below $p_2$ is dry adiabatic. 
In equilibrium $\frac{dT}{dt} \propto \frac{dF}{dp} = 0$. 
Then using the two stream approximation of radiative transfer 
\begin{equation}
\frac{dF}{dp_\uparrow} = \frac{dF}{dp_\downarrow} = \frac{\kappa}{2g}(F_\uparrow - F_\downarrow)
\end{equation}
with appropriate boundary conditions, we can analytically solve for the upward and downward heat flux at every level in the stable layer. 
Of interest for our problem is the net heat flow from the deep/stable layers to the shallow layer that is convectively coupled to the photosphere. 
This is 
\begin{equation}
F_\uparrow (p_1) - F_\downarrow (p_1) = \frac{4 F_2 g p_0 + F_1 (p_2^2 - p_1^2) \kappa_0}{4 g p_0 + (p_2^2 - p_1^2)\kappa_0} - F_1
\end{equation}
where $F_1 = F_\downarrow (p_1)$ and $F_2 = F_\uparrow (p_2)$ are the boundary conditions. 
We can solve for these boundary conditions using the uniquely determined temperature structures above and below the stable layer, i.e. 
\begin{equation}
F_1 = \frac{\kappa_0 \sigma}{g} \int_{0}^{p_1} \exp\left[\frac{\kappa_0 (p^2 - p_1^2)}{2 g p_0} \right] T(p)^4 (p/p_0) dp
\end{equation}
\begin{equation}
F_2 = \frac{\kappa_0 \sigma T_2^4}{g} \int_{p_2}^\infty \exp\left[\frac{\kappa_0 (p_2^2 - p^2)}{2 g p_0} \right] \left(p/p_2\right)^{4 R/c_p} (p/p_0) dp
\end{equation}
Using this process, for a given $(T_{\text{int}}, q_{\text{max}}) \implies (p_2, T_2)$ we solve the above nonlinear equation to obtain $T_e \implies (p_1, T_1)$ using the condition $F_\uparrow (p_1) - F_\downarrow (p_1) = \sigma (T_e^4 - f T_0^4)$. 
This defines the observed effective temperature as a function of the internal effective temperature $T_e(T_{\text{int}})$, shown as the dashed-dotted curves in Figure~\ref{fig:methane-cancel}. 
In order to relate the observed effective temperature $T_e$ to the adiabatic effective temperature $T_{ab}$, one must additionally consider the effect of latent heat, shown as solid curves in Figure~\ref{fig:methane-cancel}. 
Then the net effect, $T_e(T_{ab})$ is shown as dashed curves in Figure~\ref{fig:methane}, where $\Delta T = T_e - T_{ab}$.  
Figure~\ref{fig:PT} shows a sample temperature/pressure profile to illustrate the contributions from each source.
Similar results are shown for water in Figure~\ref{fig:water}. 
These figures show $\Delta T$ under two assumptions: an initially dry adiabatic atmospheric profile (solid curves) where moist adiabaticity is not considered, and an initially moist adiabatic upper layer (dashed curves), as described in detail in the text. 
In the dry adiabatic case, $\Tint = T_{ab}$. 
One may notice that $\Delta T = T_e - T_{ab}$ arising from the dry adiabatic case is substantially smaller than $T_e - \Tint$ in the wet adiabatic case (dashed-dotted curves in Figure~\ref{fig:methane-cancel}). 
There are two reasons for this. 
First, the dry adiabatic lapse rate is steeper than the wet pseudo-adiabatic lapse rate, which improves the efficiency of radiative transfer. 
Second, the super-adiabatic stable region makes less of a difference when the initial profile was already comparatively steeper than the moist adiabatic case. 
For this reason, the results are less sensitive to assumptions about the initial temperature profile than one might initially expect. 
 \\

Finally we must consider the deposition of sunlight. 
The primary absorber of sunlight in Uranus' troposphere is methane vapor \cite{marley-mckay1999}, while the primary absorber of infrared light is hydrogen collisions. 
At 1 bar with 2\% methane for example, accounting for methane absorption plus Rayleigh scattering vs. thermal absorption by hydrogen, the ratio between visible to thermal infrared opacity is approximately $\kappa_{\odot} / \kappa_{t} < 10^{-2}$ averaging over a broad band in wavelength. 
This value of course is not unique; there are windows at certain wavelengths, the methane abundance changes rapidly with depth, and this simple calculation neglects absorption by haze and cloud particles. 
Nevertheless $\kappa_{\odot} / \kappa_{t}$ is sufficiently small that we can plausibly argue that sunlight penetrates significantly beyond the 1 bar level, and most of the sunlight is absorbed deeper in the atmosphere. 
This can be parameterized by simply arguing some fraction $f$ of sunlight is absorbed above the cloud level, and $1-f$ is absorbed below. 
In principle this procedure accommodates any value of $f$, but for our purposes for simplicity we approximate using the limiting cases $f\rightarrow 0$ for the shallow ($\sim 1$bar) methane condensation level, while $f \rightarrow 1$ for deep ($\sim 100$bar) water clouds. 
The difference for the stable layer is that in radiative equilibrium, the flux through the stable layer must balance with $\sigma (T_e^4 - f T_0^4)$ as explained above. 

\subsection{Meteorological implications}
The equilibrium solution will be stable if the virtual potential temperature $\theta_v$ is monotonically decreasing with increasing pressure between $p_1$ and $p_2$. 
\begin{equation}
\theta_v(p, T) = T_v(p, T) \left( \frac{p_0}{p} \right)^{\nabla_{ab}}
\end{equation}
\begin{equation}
T_v(p,T) = T \left( 1 - q (1 - \epsilon) \right)^{-1}
\end{equation}
In the equilibrium cases discussed here, this condition is always satisfied (see dashed red curve in Figure~\ref{fig:PT}). 
This result should be fully general; it does not depend on, for example, the choice of atmospheric opacity. 
In order to trigger a convective instability, the upper layer would need to become more dense than the deep layer 
(note in this section we use the word ``dense'' in the virtual potential temperature sense, i.e. accounting implicitly for adiabatic expansion/compression). 
In order to accomplish this by cooling, the upper layer would need to cool such that the bottom of the upper layer satisfies $q<\qcrit$, so that further cooling makes the gas mixture more dense rather than less dense. 
However, as soon as $q$ becomes infinitesimally less than $\qcrit$, it will be more dense than the material in the stable layer directly beneath it, even while remaining less dense than the well mixed gas in the deep layer. 
This will cause a small convective instability wherein a portion of the stable layer is eroded into the upper layer, causing the stable layer to thin and restoring the upper layer to satisfy $q=\qcrit$ at a new pressure level. 
This will happen in all cases where the stable layer is of finite thickness. 
Therefore, in order to trigger a convective instability with the deep layer, the stable layer must vanish completely, reducing to a compositional discontinuity between the upper and deep layers. 
Such a scenario in the limit of heat transport by thermal conduction and an inviscid fluid results in an infinitesimal thermal boundary layer with an infinitesimal temperature discontinuity. 
In the optically thin radiative transfer case, the situation is somewhat more subtle, because rather than heat flux diverging to infinity, it converges on a finite value (in the two stream approximation: the upward heat flux from below minus the downward heat flux from above). 
Nevertheless, if there is a significant temperature difference between the upper and deep layer (as would be required in order to trigger a convective instability) this finite value is orders of magnitude larger than the luminosity of the planets in question, for any appreciable temperature discontinuity. 
This result contrasts with previous findings \cite{li-ingersoll2015}, which posited the system may behave as a relaxation oscillator when the upper layer cools sufficiently to become over-dense and trigger a convective instability. 
This previous study did not explicitly account for radiative transfer across the stable layer, instead dynamically cooling from above and treating the stable layer as a perfect insulator. 
In our case, we find instead that the profile relaxes into a state of global radiative-convective equilibrium; eventually the upper layer stops cooling as the heat it loses to space balances with the heat radiated across the stable layer. \\

After solving for the equilibrium atmospheric profile, it is possible to compute the convective available potential energy (CAPE), as well as the activation energy required to disrupt the equilibrium stable layer. 
Both quantities can be computed in the same way, using 
\begin{equation}
E = \int_{p_a}^{p_b} R_d (T_{v,f} - T_{v,i}) \frac{dp}{p}
\label{eq:CAPE}
\end{equation}
where $T_{v,f}$ is the final virtual temperature profile (for example, a moist adiabat) , $T_{v,i}$ is the initial virtual temperature profile (e.g. in radiative-convective equilibrium with a stable layer), and $p_a$ and $p_b$ are bounding pressures that satisfy $T_{v,i}=T_{v,f}$. 
The blue and red dashed curves in Figure~\ref{fig:PT} provide a visual example for how this calculation can be done. 
This computed activation energy in units of energy per mass, can be converted into updraft velocities required to disrupt the stable layer. 
In the case of Saturn, if $\qcrit \sim \qmax$ as suggested in \cite{li-ingersoll2015}, then the activation energy required to disrupt this stable equilibrium is quite small, requiring updraft velocities of only a few meters per second. 
Therefore the basic premise of \cite{li-ingersoll2015} can still be valid, although it may require some additional mechanism to jump-start the process, for example a strong updraft or entrainment erosion as described in \cite{friedson-gonzales2017}. 
In the case of Uranus and Neptune with methane mixing ratios further from the critical value, the activation energy necessary to initiate a convective instability is correspondingly larger, requiring updraft velocities of order tens of meters per second, far larger than expected convective velocities. 
However, if such a disruption were able to occur by some anomalous updraft, the resulting storm would be extremely energetic, with CAPE exceeding 10 J/g, substantially larger than even the most extreme weather events on Earth. 

\subsection{Interior implications}
\label{interior}
\begin{figure}
\centering
\includegraphics[scale=.35]{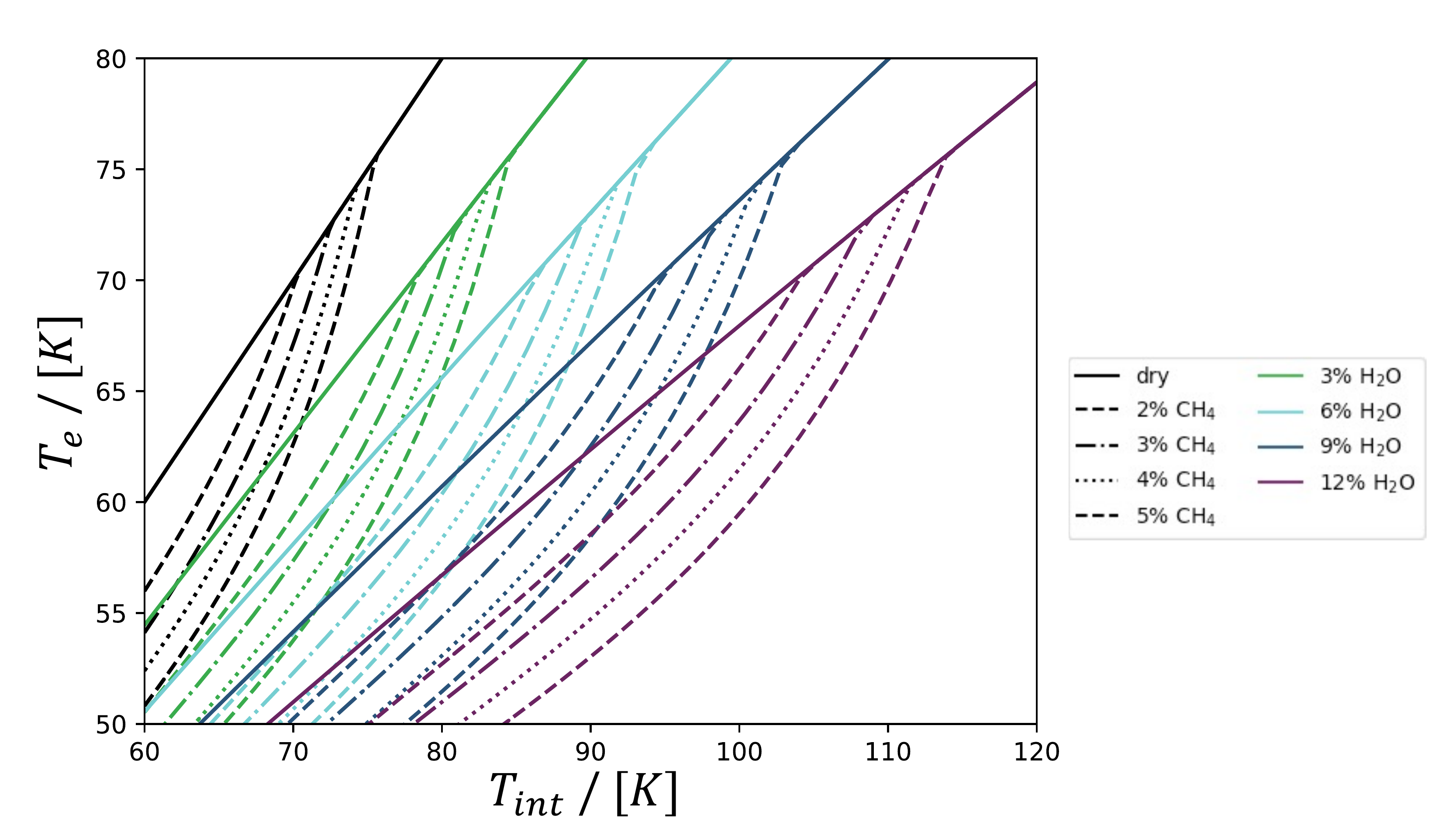}
\caption{Apparent effective temperature $T_e$ as a function of internal equivalent effective temperature $\Tint$ considering both methane and water. 
Colors correspond to water abundance, increasing downward from 0\% to 12\% water. 
The curve style corresponds to methane abundance, increasing downward from 0\% to 5\% abundance. 
}
\label{fig:Tes}
\end{figure}
Interior models of giant planets generally assume a 1-bar equivalent temperature that sets the internal entropy of the planet. 
This is usually done using the atmospheric temperature, and corrected for any non-adiabatic behavior in the atmosphere. 
Non-adiabatic models for Uranus' and Neptune's interiors and thermal evolution have recently been carried out (e.g. \cite{nettelmann+2016}, \cite{vazan-helled2020}, \cite{scheibe+2021}), finding self-consistent solutions to Uranus' contemporary heat flow. 
Our results are still relevant to many of these models. 
The so called thermal boundary layers at depth from \cite{nettelmann+2016} and \cite{vazan-helled2020} retain an adiabatic convective envelope set by a 1-bar equivalent temperature. 
Furthermore, the U-1 and U-2 models from \cite{vazan-helled2020} involve a convective envelope of homogeneous composition, whose temperature profile is also set using a 1-bar equivalent temperature. 
Although our results make no direct statement about the behavior of the deep interior of the planet that may include extended regions of static stability, it still applies to the thermal evolution of these upper envelopes. 
Because Uranus and Neptune possess magnetic fields, and the latter possesses a substantial internal heat flow, we deduce that both must be convective at depth, and that this adiabatic description is probably relevant for at least some fraction of the interior of Uranus and Neptune. 
The primary source of non-adiabatic behavior in the atmosphere is condensation, usually accounted for by the sub-adiabatic gradients caused by latent heat of condensation. 
In Figure~\ref{fig:Tes}, we show the apparent effective temperature $T_e$ against the internal equivalent effective temperature $\Tint$ for various envelope condensate abundances. 
If convective inhibition does occur, it will largely cancel out the effect of moist adiabaticity. 
Therefore, using a dry adiabat to guess the 1-bar equivalent temperature is a better approximation than accounting for latent heat alone but neglecting convective inhibition. 
A better approach could be to use the analytic scaling relationships in Section~2 to estimate the magnitude of these effects. 
The best approach would be to explicitly model the effect of convective inhibition, using methods from this work or \cite{leconte+2017}. 


\section{Evolutionary model}
\label{evolution}
We present an adiabatic thermal evolution model of Uranus and Neptune. 
As discussed, treating the interior as adiabatic is probably inappropriate for the ice giants \cite{helled+2020}. 
Nevertheless it provides a convenient framework to understand the effect of methane condensation on these planets' thermal histories in the absence of an accepted interior model. 
For an adiabatic model, we assume the total heat content of the planet's interior to be a linear function of its adiabatic equivalent effective temperature $T_{ab}$. 
\begin{equation}
\int_{M_{\text{min}}}^M c_p T dm = A \bar{c_p} M T_{ab}
\end{equation}
One way of thinking about this equation is to imagine a small set of layers, or possibly even one layer, in the form of concentric shells, each of which is isentropic and homogeneous but of different composition to neighboring layers, with negligible thermal boundary layers between them as would be fluid dynamically expected for a low viscosity system. 
Beneath this set of shells there could be a region, possibly a substantial fraction of the planet, where there is a compositional gradient and therefore inefficient convective transport. 
This deeper region would not contribute to $A$ or to the resulting thermal evolution of the planet because it stores primordial heat. 
In reality, there would be non-zero thermal diffusion from a stably stratified interior portion of the planet and its convective envelope, slowing down planetary cooling. 
However, if the diffusion timescale for the planet is longer than the age of the solar system, then this contribution would be small. 
It is not guaranteed that the diffusion timescale is in fact longer than the age of the solar system. 
It depends on the (unknown) thermal transport properties of the (unknown) compositional constituents of the ice giants' interiors, and is further complicated by the possibility of thermal transport by double diffusive convection. 
Therefore we acknowledge this description of the interior evolution is imperfect, but it does at least approximately describe a wide variety of possible interior behaviors, and provides a convenient framework to self-consistently assess the relative influence of convective inhibition on thermal evolution while remaining agnostic about the details of the ice giants' interior structures. 
A full description of Uranus' and Neptune's thermal evolutions would require a detailed interior model. 
Nevertheless under our assumptions, the parameter $A$ is approximately constant through time because the Gruneisen parameter is rather insensitive to temperature, and its value is set by the fraction of the total mass that is fully convective. 
Some fraction of the planets must be convective in order to generate their observed magnetic fields. 

The rate of cooling depends on the apparent effective temperature $T_e$, while the heat content of the bulk of the interior is linearly related to the adiabatic equivalent effetive temperature $T_{ab}$.  
The equation governing thermal evolution is then
\begin{equation}
\label{eq:evol}
4 \pi R^2 \sigma (T_e^4 - T_0^4) = - A \bar{c_p} M \frac{d T_{ab}}{dt}
\end{equation}
where $T_0$ is the equilibrium effective temperature of the planet with sunlight if there were no internal heat source. 
The radius $R$ is treated as constant because we are concerned with most of the evolution where the body is degenerate, not any early very hot phase. 
The steady increase in solar luminosity (i.e. time variation of $T_0$) is ignored. 
In order to solve the thermal evolution equation, we need an explicit relationship between $T_e$ and $T_{ab}$. 
This is done using the method from Section~\ref{atmosphere}, with the results for methane in Figure~\ref{fig:methane} and for water in Figure~\ref{fig:water}. 
In general the relationship between $T_e$ and $T_{ab}$ depends on the condensate bulk interior abundance $\qmax$. 
Without condensation, the relationship is $T_e = T_{ab}$. 
In this case, Equation~\ref{eq:evol} can be non-dimensionalized into the canonical form
\begin{equation}
\label{eq:nocond}
\frac{x^4 - x_0^4}{1 - x_0^4} = - \tau_K \frac{dx}{dt}
\end{equation}
where $x \equiv T_e / T_e^{(\emptyset)}$ where $T_e^{(\emptyset)}$ is the apparent effective temperature today, and $x_0 = T_0 / T_e^{(\emptyset)}$. 
The Kelvin timescale $\tau_K = \frac{A \bar{c_p} M T_e^{(\emptyset)}}{4\pi R^2 \sigma} (T_e^{(\emptyset)4}-T_0^4)^{-1}$ scales how long it takes to cool to $T_e^{(\emptyset)}$ from an initial arbitrarily hot state. 
In the asymptotic case $T_e^{(\emptyset)} \gg T_0$ (not true for Uranus!), this is about $\tau_K/4$, but can be a different fraction of $\tau_K$ in general. \\

Accounting for condensation, Equation~\ref{eq:evol} becomes 
\begin{equation}
\label{eq:cond}
\frac{dx}{dx_{ab}} \frac{x^4 - x_0^4}{1 - x_0^4} = - \tau_K \frac{dx}{dt}
\end{equation}
where $x_{ab} = T_{ab} / T_e^{(\emptyset)}$. 
The difference between Equations~\ref{eq:nocond}~and~\ref{eq:cond} then straightforwardly demonstrates the effect of convective inhibition by condensation on the planets' thermal evolution: it alters the rate of cooling by a factor of $\frac{dx}{dx_{ab}}$. 
As we will see, this factor can be greater than or less than unity.  This means the effect can either speed up or slow down the rate of change of the planets' apparent effective temperature at different points in its thermal history. 
This is especially important for understanding the results for water. \\

\begin{figure}
\begin{subfigure}{.5\textwidth}
	\centering
	\includegraphics[scale=.35]{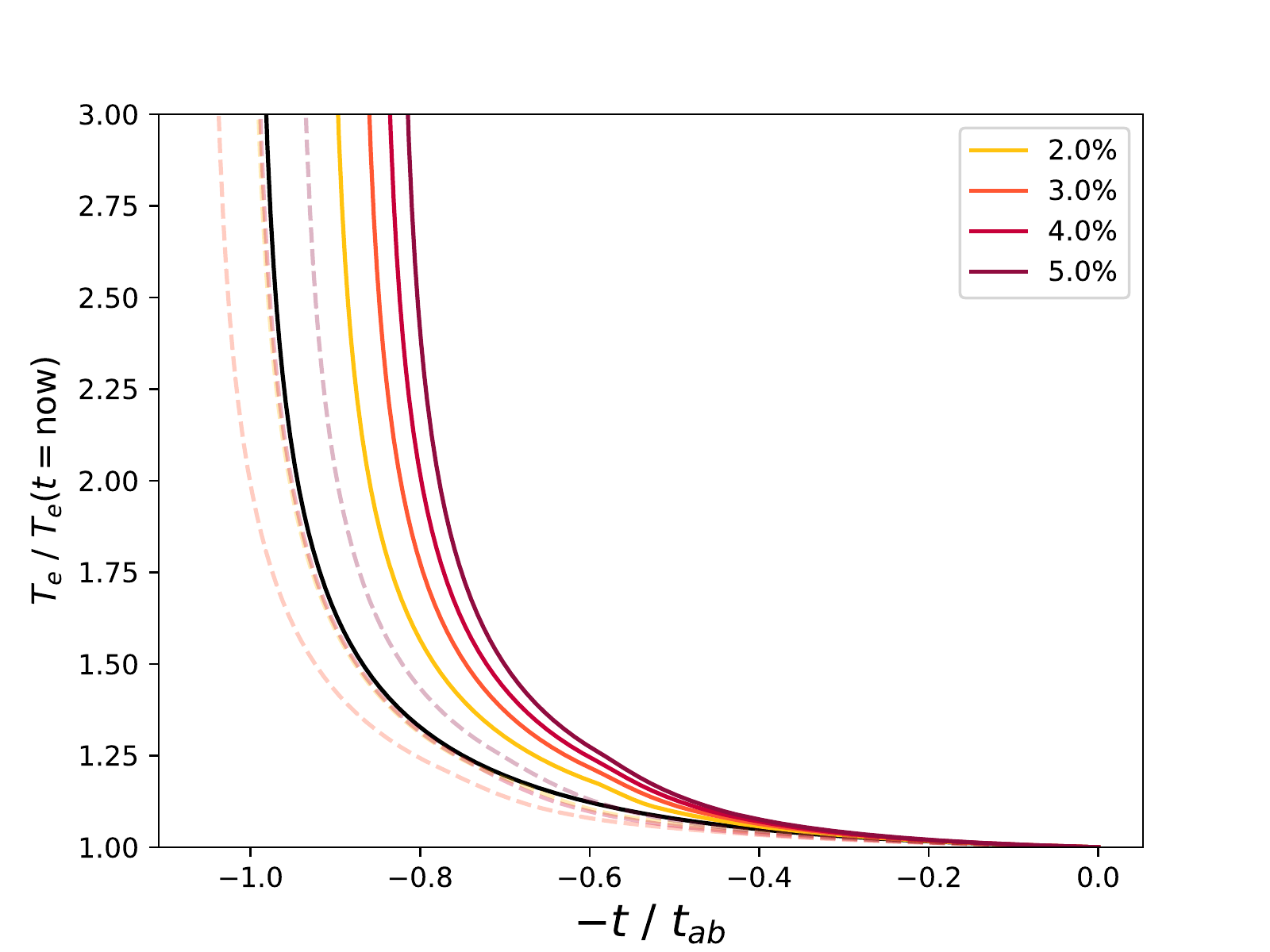}
	\caption{Uranus thermal evolution}
	\label{fig:ura-meth-evol}
\end{subfigure}%
\begin{subfigure}{.5\textwidth}
	\centering
	\includegraphics[scale=.35]{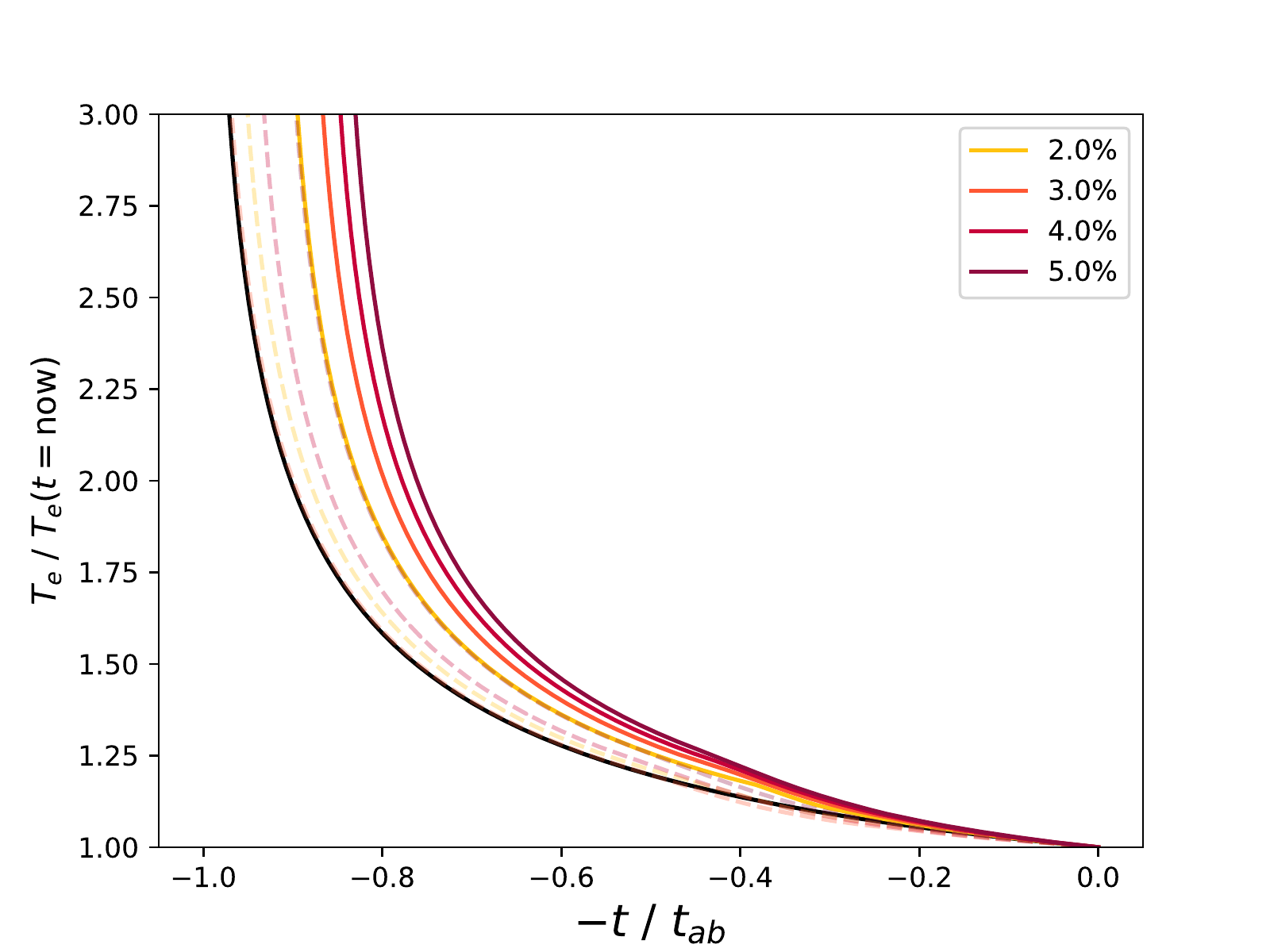}
	\caption{Neptune thermal evolution}
	\label{fig:nep-meth-evol}
\end{subfigure}%
\caption{Thermal evolution model for Uranus and Neptune, with different colored curves representing different methane abundances. 
The x-axis is the time before the present day, scaled to the cooldown time in the dry adiabatic (no condensation) case. 
The y-axis is $x$ from Equation~\ref{eq:cond}.
Line styles and colors are identical to Figure~\ref{fig:methane}.}
\label{fig:meth-evol}
\end{figure}

\begin{figure}
\begin{subfigure}{.5\textwidth}
	\centering
	\includegraphics[scale=.35]{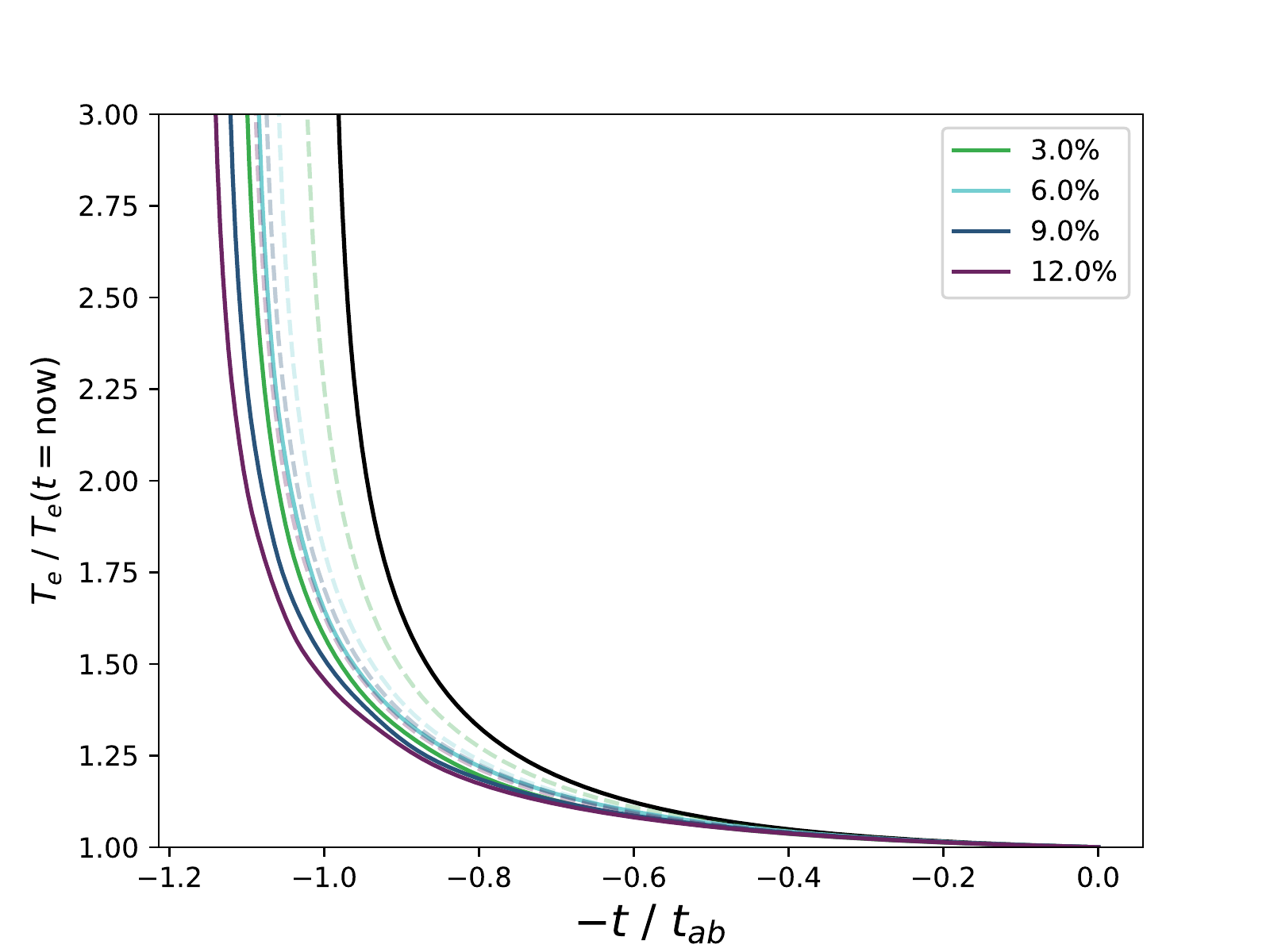}
	\caption{Uranus thermal evolution}
	\label{fig:ura-water-evol}
\end{subfigure}%
\begin{subfigure}{.5\textwidth}
	\centering
	\includegraphics[scale=.35]{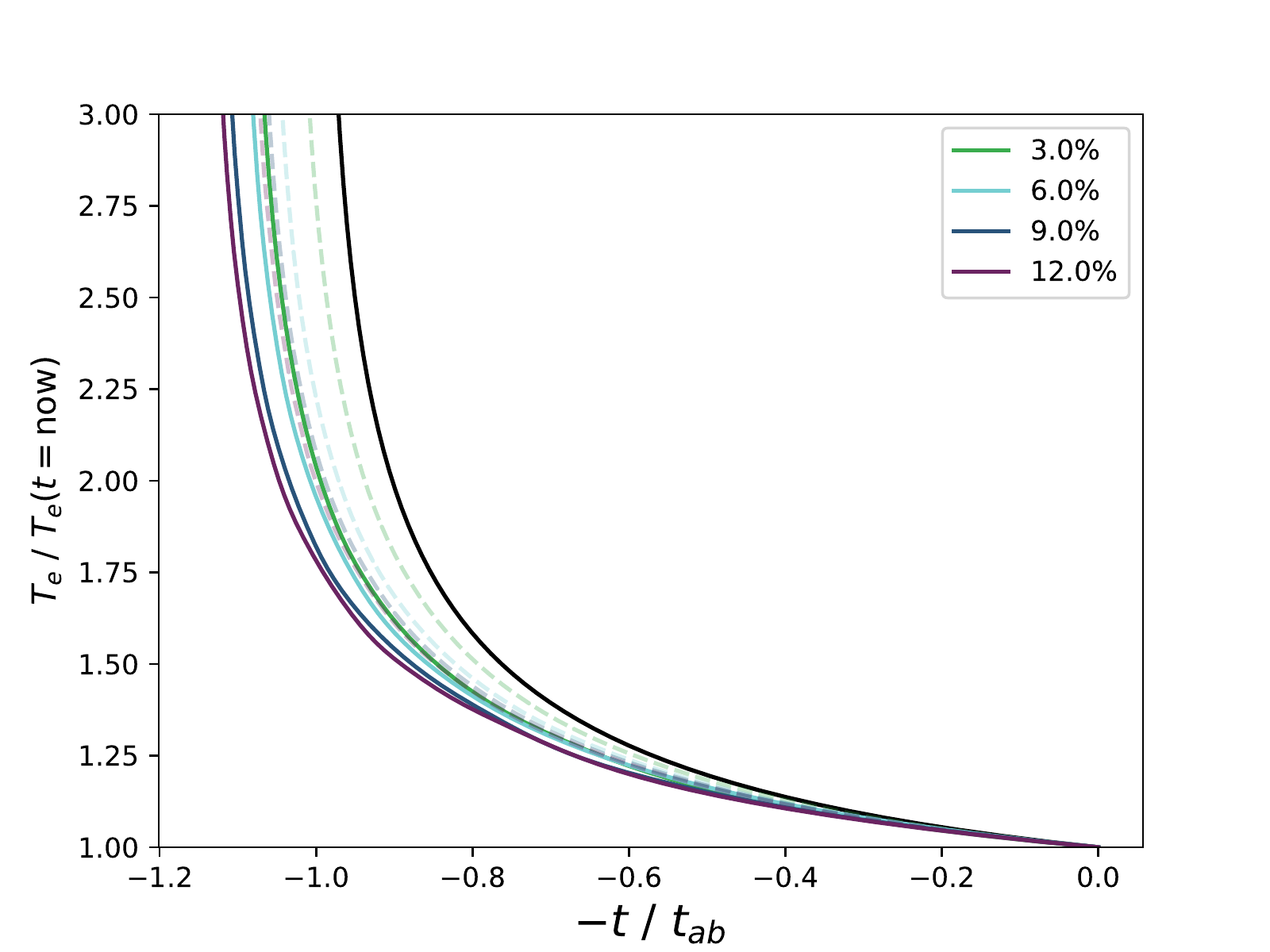}
	\caption{Neptune thermal evolution}
	\label{fig:nep-water-evol}
\end{subfigure}%
\caption{Thermal evolution model for Uranus and Neptune, with different colored curves representing different water abundances. 
Axes are identical to Figure~\ref{fig:meth-evol}. 
Line styles and colors are identical to Figure~\ref{fig:water}.}
\label{fig:water-evol}
\end{figure}

The fact that Equation~\ref{eq:cond} retains the Kelvin timescale $\tau_K$ makes this formulation especially convenient. 
This allows us to directly compare the fraction of that timescale that a given evolutionary model takes to cool from arbitrarily hot bodies to their current temperatures for different assumptions of the condensate abundance $\qmax$. 
Leaving the the effect in terms of the Kelvin timescale allows our results to be roughly independent of accurate interior models, because $\tau_K$ implicitly encodes an arbitrary interior model. 
The results for methane are shown in Figure~\ref{fig:meth-evol}, and for water in Figure~\ref{fig:water-evol}. 
The results for methane are relatively straightforward; for the early stages of Uranus and Neptune's thermal histories, the effect of methane is unimportant, because the atmosphere is warm enough that methane does not condense anywhere. 
As the atmosphere cools, methane begins to condense, at first in the stratosphere above the radiative convective boundary.  
As cooling continues, convective inhibition begins to extend the radiative-convective boundary downward, as superadiabatic gradients can be stable. 
At this point, the measured effective temperature $T_e$ departs from its adiabatic equivalent $T_{ab}$, causing the effective temperature to drop faster than the interior is cooling. 
As cooling continues, the layered system described in Section~\ref{atmosphere} emerges, and perhaps persists today \cite{guillot1995}. \\

The case of water, shown in Figure~\ref{fig:water-evol}, the behavior is more subtle. 
In this case, thermal evolution is actually slowed down compared to the adiabatic case. 
This contrasts with previous findings \cite{kurosaki-ikoma2017}, which considered the effect of moist adiabaticity (i.e., latent heat) but did not quantify the effects of convective inhibition. 
If we consider latent heat only, we obtain results in good agreement with this previous study. 
In their case, thermal evolution is sped up, because the atmosphere initially remains warm while the interior cools. 
This allows the planet to lose heat efficiently when condensation first occurs, speeding up evolution. 
Our findings demonstrate that the effect of convective inhibition overwhelms the effect of moist adiabaticity, so that our story is the opposite. 
Early on, as condensation occurs, we find the atmosphere cools faster than the interior, reducing cooling efficiency. 
Later on, $\frac{dx}{dx_{ab}}$ from Equation~\ref{eq:cond} becomes less than unity, as demonstrated by the negative slope at low temperatures in Figure~\ref{fig:water}. 
Therefore in the case of water, condensation early in the ice giants' thermal histories caused the effective temperature to drop faster than the internal temperature, analogous to what happened with methane condensation more recently. 
However, this temporary speedup of $\frac{d T_e}{dt}$ coincides with a loss of luminosity, slowing down the rate at which the interior loses heat. 
Then, over subsequent evolution, the interior cools inefficiently, and in recent history the effective temperature changes \textit{slower} than the internal temperature. 
The net effect is a cooldown time that is longer than the dry adiabatic case. \\

Because the water and methane cloud decks are well separated, the superposition of the two effects is straightforward.  
Immediately beneath the methane cloud deck, the behavior can be accurately modeled as a dry adiabat, because the water mixing ratio is so small at these relatively low temperatures. 
Therefore the results of modeling the whole atmosphere with both cloud decks explicitly is virtually identical to using $\Delta T_{tot} = \Delta T_{CH_4} + \Delta T_{H_2 O}$ from Figures~\ref{fig:methane} and \ref{fig:water}. 

\section{Discussion}
\label{discussion}
Provided $\qmax > q_{\text{crit}}$, when the planet cools to a temperature low enough for condensation,  convection can be interrupted. 
At this point the apparent effective temperature departs from the internal equivalent effective temperature, by the mechanism described in Section~\ref{atmosphere}. 
We can solve for the equilibrium configuration to derive the apparent effective temperature $T_e$ as a function of $\qmax$ and $\Tint$. 
These results are shown as the dashed-dotted lines for methane in Figure~\ref{fig:methane}, and Figure~\ref{fig:water} for water. 
Interior  modelers should bear in mind that the internal 1-bar equivalent temperature may depart from simple adiabatic extrapolation of the troposphere by nearly a significant factor (see Section~\ref{interior} and Figure~\ref{fig:Tes}). \\

This behavior leads to the evolutionary behavior observed in Figure~\ref{fig:meth-evol}. 
Before condensation occurs, the planet cools normally. 
Upon the onset cloud formation, the apparent effective temperature drops rapidly. 
However, upon reaching the minimum, the apparent effective temperature actually begins to decrease more slowly than the fully adiabatic case. 
This effect is present in Figure~\ref{fig:methane} but is more apparent in Figure~\ref{fig:water}. 
The net effect for the ice giants is a net speedup of thermal evolution for methane, and a net slowdown for water. 
The magnitude of this speedup or slow down can be no more than 15\% in either case, assuming 5\% methane molar abundance or 12\% water molar abundance. 
Both effects can occur simultaneously, and superimpose straightforwardly because their cloud decks are well separated. \\

As Neptune continues to cool, methane will begin to behave similarly to water, exhibiting a local minimum in $\Delta T(T_{ab})$. 
This local minimum is seen most clearly for water in Figure~\ref{fig:water} around $T_{ab}=$180K, but can also be seen for the dashed curves in Figure~\ref{fig:methane} around 75K, and would be present in the solid curves if the x-axis extended to lower temperatures. 
As the planet continues to cool below this local minimum, the slope of $\Delta T(T_{ab})$ becomes negative, and the rate of change of the thermal state of the atmosphere slows. 
Consider the implications of this for methane clouds near the 1-bar level. 
This state persists for longer from a thermal evolution perspective than an arbitrary/random thermal state. 
That is, this state is a local minimum in $\Delta T(T_{ab})$, meaning the planet reaches this state faster than it would if it were cooling adiabatically, and leaves this state more slowly than it would if it were cooling adiabatically. 
Therefore these planets will spend a longer portion of their thermal histories in the state where the cloud level is $\sim 1-10$bars than they would in a thermal evolution model that does not consider convective inhibition by condensation. 
Perhaps this consideration renders the surprising similarity of Uranus and Neptunes' atmospheres' shallow temperature structures despite their vast difference in insolation somewhat less improbable than it first appears. \\

We must consider whether this atmospheric structure is compatible with existing data, especially Voyager radio refractivity data. 
The current data has been shown to be consistent with many different models, including subadiabatic, adiabatic, moist adiabatic, and superadiabatic temperature gradients \cite{helled+2020}. 
The data has also been shown to be compatible with a wide range of temperature structures and methane abundances \cite{lindal+1987} \cite{lindal1992}. 
The data itself shows a layer of rapidly varying refractivity near the condensation level, generally interpreted to be methane clouds \cite{lindal+1987} \cite{lindal1992}\cite{marley-mckay1999}. 
Another interpretation of the same data supports a layer of superadiabatic temperature lapse rate in the cloud-forming regions of these planets \cite{guillot1995}. 
In general, our understanding of the thermal structure of the ice giant atmospheres is incomplete, as the results from Voyager 2 refractivity data are model dependent, with a particular degeneracy between assumed methane enrichment and temperature structure. 
In order to disentangle these variables and have a more confident understanding of these planets' atmospheres' thermal structures, we must return with a mission. 
It should be a priority for a future mission to independently measure methane abundance and temperature, perhaps with entry probes or a well designed microwave radiometer experiment. 
\\

These general findings do not consider the long term stability of stable layers in the atmosphere. 
As long as the stability timescale is greater than the relaxation timescale for a stable layer, the results should approximately reflect reality. 
However, the stability timescale is poorly constrained \cite{friedson-gonzales2017}. 
If it is sufficiently short, this could further complicate the dynamics.
If that condition is satisfied, then even if stable layers are intermittently interrupted by massive internal plumes, large meteor impacts, or instability due to long term erosion by entrainment, they will reform again on geologically short timescales ($\sim 100$yr). 
Therefore the thermal evolution will be governed primarily by the equilibrium state, and not possible intermittent periods of enhanced activity. 
Intermediate states where the equlibrium configuration is thinned over time but not totally destroyed by entrainment erosion would in general reduce the magnitude of $\Delta T(\Tint)$, so the findings in this paper should be considered an upper bound. 
Furthermore, we use a highly simplified thermal evolution model, not considering changes in planetary radius or explicitly accounting for the effects of non-adiabaticity at depth. 
Seasonal variations in insolation were not included in the model, as these variations average out over geologic time. 
However, seasonal variations have been shown to create local temperature variations of order 10K \cite{orton+2007}, comparable to the magnitude of the effect of convective inhibition by methane. 
The possible dynamical and evolutionary consequences could be the subject of future work. 
Our atmospheric model also did not explicitly include the condensate opacities, and may therefore not capture possible feedback mechanisms. 
We discuss further the possible effects of opacity variation due to condensation in the following paragraphs. 
For these reasons, this work should be considered exploratory, and further work is needed in order to more confidently establish the thermal histories of the ice giants while accounting for convective inhibition. \\

Here we must include a discussion about the effects of opacity variation due to condensation, which are not considered in this model but which are certainly important for a fully complete understanding of Uranus and Neptune's thermal states and thermal histories, and has been considered explicitly by prior works, e.g. \cite{kurosaki-ikoma2017}. 
The variation of opacity affects our results in two important ways: first, by changing the radiative-convective boundary as vapor condenses out of the stratosphere; and second, by affecting radiative transfer within and across the layer of stable stratification caused by convective inhibition. \\

We begin by discussing the stratospheric effect of opacity variations due to condensation. 
Water and methane are both more opaque than hydrogen in the thermal infrared, therefore as the planet cools and these volatiles begin to condense and rain out of the stratosphere, the stratosphere becomes more transparent and the radiative-convective boundary deepens. 
At fixed effective temperature, the temperature at the radiative-convective boundary is relatively unchanged, therefore decreasing the opacity of the stratosphere has the net effect of decreasing the entropy of the troposphere at fixed effective temperature. 
Therefore during this stage, the temperature of the troposphere is cooling faster than the effective temperature of the planet as the stratosphere extends downward. 
Convective inhibition only begins to become relevant when the stratosphere has cooled sufficiently such that the radiative-convective boundary has a lower vapor mixing ratio than the bulk abundance. 
By the time this occurs, the bulk of the stratosphere is cooler than the radiative convective boundary by approximately a factor of $2^{1/4}$, and therefore relatively dry due to the highly sensitive dependence of saturation vapor pressure on temperature. 
So the important stratospheric effect due to opacity variation we have just described qualitatively, essentially predates the onset of convective inhibition. 
This allows us to neglect these dynamics in our context, although we caution the reader that a fully realistic consideration of the effects of condensation must also include the effects of opacity variations, which are important. \\

The second effect of opacity variation is on the radiative transfer across the stable layer. 
To estimate the importance of this effect, we modifed our method so that we increased the opacity of the deep layer and and stable layer (see Figure~\ref{fig:sketch}) by an order of magnitude. 
This changes our results for contemporary methane clouds by no more than 2\%, and for contemporary deep water clouds by $<$0.01\%. 
The effect for water clouds is larger earlier in its evolution, but is always $<$2\%. 
In either case, the direction of this consideration is to increase the magnitude of $\Delta T$. 
The reason this matters less for deep clouds is because the opacity is already very large, and the effect due to deep clouds is accurately approximated by Equation~\ref{eq:approx}, which assumes a high opacity limit. 
For shallower clouds where the details of thermal transport are more relevant, it affects the results, but only as a relatively small correction even assuming a very large order of magnitude change in opacity.  \\

If there are indeed layers of static stability in the troposphere or deep atmosphere of Uranus and/or Neptune, then they should support gravity waves. 
Whether we expect gravity waves to be excited, what their general characteristic would be, and whether they could be detected from space (for example using an Doppler imager) is a subject worthy of future theoretical consideration. \\

Whatever the uncertainties about the specifics, the basic physical mechanism is likely to be important in the ice giants because of their highly enriched atmospheres. 
There may be additional stable layers, for example a silicate cloud level beneath the water cloud level, or a sulfide/ammonia cloud level. 
We focus on only two in this work to demonstrate the general principle without getting bogged down in largely unconstrained assumptions about the envelope enrichment in each species. 
However, the intuition we build here for methane and water can be straightforwardly applied to other cloud levels using exactly the same method. 
This method is also likely to be applicable to the majority of exoplanets, ranging from super-Earths and water worlds with hydrogen envelopes, to metal-enriched gas giants. 
It is clear from this work that thermal evolution and internal thermal structure may be profoundly influenced by convective inhibition by condensation. 
Any complete model of thermal evolution or internal structure is advised to consider convective inhibition.

\subsubsection*{Acknowledgments}
This work has been funded by the NASA FINESST program. This work has greatly benefited from the ISSI Ice Giants Science Team meetings of 2019 and 2020 in Bern.  We would also like to thank the anonymous reviewers for their invaluable feedback that greatly strengthened this work.

\bibliography{library}{}
\bibliographystyle{apalike}

\end{document}